\documentclass[10pt,aps,pre,twocolumn,amsmath,amssymb]{revtex4-1}

\usepackage{amssymb}
\usepackage{amsmath}
\usepackage{graphicx}

\def\Eq#1{Eq.~(\ref{#1})}
\def\Eqs#1{Eqs.~(\ref{#1})}
\def\Fig#1{Fig.~\ref{#1}}
\def\Figs#1{Fig.~\ref{#1}}
\def\nd{\noindent}
\def\nm{\nonumber\\}

\def\>{\rangle}
\def\<{\langle}

\def\tr{\text{Tr}}
\def\dg{\dagger}
\def\adj{\text{adj}}
\def\lam{\lambda}

\def\w{\omega}
\def\del{\delta}
\def\gam{\gamma}
\def\s{\sigma}

\def\al{\alpha}
\def\bt{\beta}
\def\th{\theta}	
\def\eps{\epsilon}
\def\d{\partial}
\def\T{\text{T}}
\def\i{\text{i}}
\def\t{\text{t}}
\def\st{\text{st}}
\def\amp{\text{amp}}
\def\ph{\text{ph}}
\def\u#1{\underline{#1}}
\def\ud#1{\underline{\underline{#1}}}
\def\hhat#1{\hat{\hat{#1}}}

\begin{document}

\title{General symmetry in the reduced dynamics of two-level system}

\author{B.~A.~Tay }
\email{BuangAnn.Tay@nottingham.edu.my}
\affiliation{Department of Foundation Studies, Faculty of Science and Engineering, The University of Nottingham Malaysia Campus, Jalan Broga, 43500 Semenyih, Selangor, Malaysia}

\date{\today}

\begin{abstract}
We study general transformation on the density matrix of two-level system that keeps the expectation value of observable invariant. We introduce a set of generators that yields hermiticity and trace preserving general transformation which casts the transformation into simple form. The general transformation is in general not factorized and not completely positive. Consequently, either the parameter of transformation or the density matrix it acts on needs to be restricted. It can transform the system in the forward and backward direction with regard to its parameter, not as a semigroup in the time translation symmetry of dynamical maps. The general transformation can rotate the Bloch vector circularly or hyperbolically, dilate it or translate it. We apply the general transformation to study the general symmetry of amplitude damping and phase damping in two-level system. We generalize the generators to higher level systems.
\end{abstract}

%\begin{keyword}
%Symmetry \sep General symmetry \sep Density matrix \sep Open quantum systems
%\PACS 05.70.Ln
%\end{keyword}i

\maketitle

%%%%%%%%%%%%%%%%%%%%%%%%%%
%       Section          %
%%%%%%%%%%%%%%%%%%%%%%%%%%
\section{Introduction}

Quantum mechanics was originally formulated in terms of state vector \cite{Dirac} describing reversible dynamics in isolated quantum systems. As it is difficult to maintain truly isolated systems, most of the time we are dealing with systems in contact with an environment. As a result, even if a system starts off in a pure state, it would eventually evolve into a mixed state. This process cannot be fully described in terms of state vector. It is then necessary to resort to the density matrix formulation \cite{vonNeumann} for a description of the system of interest by a reduced dynamics \cite{Breuer}.

In view of the importance of the density matrix formulation in describing open quantum systems, it is worthwhile to explore aspects in which the formulation in density matrix could provide different perspectives from the formulation in state vector. The allowed transformation that keeps the expectation values of physical observables invariant serves as an interesting example that has not been paid much attention to. While in ordinary quantum mechanics the expectation values of observables are invariant under ordinary unitary transformation (or anti-unitary for time reversal) \cite{Wigner59}, in density matrix space in which the expectation values are defined by tracing the observables over the density matrices, more general transformation that keeps the expectation values invariant are in principle permitted \cite{Weinberg14}.

The possible existence of general symmetry in the density matrix space that finds no counterpart in the state vector space were discussed from two perspectives Refs.~\cite{Brink02,Tay17}. On the one hand, there is an increase of freedom in describing quantum systems when we extend quantum mechanics from the state vector space to the density matrix (Liouville) space \cite{Brink02}, which leads to degeneracy in the eigenvalues of the Liouville operator. On the other hand, reduced dynamics (irreversible) obtained by averaging out the environmental degrees of freedom from the full dynamics (reversible) in principle needs not have the same form of symmetry as the original dynamics \cite{Tay17}.

In Ref.~\cite{Weinberg14} it was also pointed out that most studies on the general transformation in irreversible systems focused on the time translation symmetry of the dynamical maps as a semigroup (dynamical aspects), whereas studies on general transformation that could transform the system forward or backward in the transformation parameter (kinematical aspects) were handful.

From the dynamical aspects, the general structure of dynamical maps as completely positive maps was studied in Ref.~\cite{Sudarshan61,Kraus83}. The form of the completely positive maps for finite-level \cite{Kossa76,Gorini78} and continuous systems \cite{Lindblad76} were also known. The parametrization of completely positive maps was discussed in Ref.~\cite{Sudarshan02}. The structure and parametrization of not completely positive maps was later understood in Ref.~\cite{Sudarshan03}, see also Ref.~\cite{Zyczkowski04}. The group structure of dynamical maps was analyzed in Ref.~\cite{Sudarshan06}. There were debates on the physical relevance of not completely positive maps \cite{Pechukas94,Shaji05}. It was found that not completely positive maps could arise from the interactions between system and environment that are initially entangled \cite{Shaji04}.

For single qubits, the structure of the completely positive trace-preserving maps was clarified in Ref.~\cite{Ruskai02}, and its actions on the Bloch ball was analyzed in Refs.~\cite{Ruskai01,Zyczkowski04,Oi01,Wodkiewicz03}. The correspondence between qubits quantum operations with special relativity was identified in Ref.~\cite{Arrighi03}.

From the kinematical aspects, the structure of general transformation that keeps the expectation values of observables invariant was analyzed in component form in Ref.~\cite{Weinberg14}. A unitarily similar symmetry of completely positive maps that keeps the spectra of dynamical matrices invariant was studied in Ref.~\cite{Zyczkowski04}. There was a closely related result on the existence of the so-called pseudo-unitary freedom in not completely positive maps \cite{Byrd10}. Symmetry in the generators of reduced dynamics in continuous system was discussed in Refs.~\cite{Tay07,Tay17}.

The conditions satisfied by the general transformation were given in a general component form in Ref.~\cite{Sudarshan61,Weinberg14}. General transformation is in general not completely positive. The negative problem can be avoided by introducing the slippage of initial conditions \cite{Suarez92,Gaspard99}, or by limiting the domain of the dynamical maps to a special set of density matrices \cite{Shaji04}.

In this paper, we focus on the kinematical aspect of general transformation by constructing generators that produce hermiticity and trace preserving transformation. Whereas most studies are carried out on the level of transformation (maps), we start from the level of generators. Furthermore, we work with the explicit superoperator form in contrast to the general component form in previous studies \cite{Sudarshan61,Weinberg14}, so that the method of Lie group can be applied \cite{Gilmore}. In this way the structure of the general transformation is clarified. In the two-level system, we work out all the possible general transformations to reveal their simple forms, and to clarify their effects on the Bloch vector.

We are able to achieve the aims by introducing a complete set of generators that produce hermiticity and trace preserving transformation. Moreover, the generators can be generalized to higher level systems. We find that all general transformations in finite-level systems, apart from the ordinary transformations, are not factorized and in general are not completely positive. However, they can be made positive by restricting the range of their transformation parameters.

We organize our discussions as follows. We first rewrite the hermiticity and trace preserving condition for the general transformation on density matrix that keep the expectation values of observables invariant in Sec.~\ref{SecSym}. They yield corresponding conditions on the generators obtained in the same scetion. In Sec.~\ref{Sec2Lev}, we realize the generators in two-level system in superoperator form, clarify the geometrical meaning of the transformations they generate and compare our results to others' works. We then apply the transformation to the study of the symmetry of amplitude damping and phase damping in two-level system in Sec.~\ref{SecInv}. In Sec.~\ref{SecNLevel}, we generalize the generators to $N$-level systems. We summarize our works in the conclusions.

%%%%%%%%%%%%%%%%%%%%%%%%%%
%       Section          %
%%%%%%%%%%%%%%%%%%%%%%%%%%i
\section{Symmetry conditions}
\label{SecSym}

The expectation value of an observable $a$ on a quantum system described by the density matrix $\rho$ is defined by $\<a\>_\rho\equiv \tr(a\rho)$. The conditions under which a general transformation will preserve the expectation value and the properties of density matrices were given in component form in Ref.~\cite{Weinberg14} and Refs.~\cite{Sudarshan61,Weinberg14}, respectively. We begin our discussion by rewriting the conditions in the superoperator language \cite{Prigogine73}.

We denote $N\times N$ matrices by small letters, such as $a, b$, and superoperators by capital letters, such as $A\equiv a\times b$. $A$ can also be written as a tensor product $a\otimes b^\t$ \cite{Zyczkowski04}, where $\t$ denotes matrix transpose. It acts on density matrix from the left as $A\rho=a\rho b$. For the convenience of later discussions, we first introduce a few operations on superoperators \cite{Prigogine73}.

\vspace{6pt}

\nd \textit{Transposition} $(\T)$. A superoperator that acts on density matrix from the right is related to a superoperator that acts on density matrix from the left, and vice versa, through transposition $\rho A=A^\T \rho$, where transposition on $A$ is defined as $(\mu A)^\T\equiv \mu b\times a$, in which $\mu$ is a complex number. We obtain this through the identity $\tr\big( (A^\T c) d\big)=\tr\big( (c A) d\big)= \tr\big( c (Ad)\big)$, and use the invariance of trace under cyclic permutations of operators. This operation is equivalent to the combined action of the `swap' and `transposition' operation defined in Ref.~\cite{Zyczkowski04}. Transposition acts on the product of two superoperators as $(AB)^\T=B^\T A^\T$.

\vspace{6pt}

\nd \textit{Adjunction} $(\dg)$. Adjunction acts on $A$ as $(\mu A)^\dg\equiv\mu^* a^\dg\times b^\dg$, where $\dg$ denotes hermitian conjugate. When it acts on the product of superoperators, we have $(AB)^\dg = B^\dg A^\dg$.

\vspace{6pt}

\nd \textit{Association} $(\sim)$. Association is equivalent to two successive operations of transposition and adjunction, which commute among themselves, $(\mu A)^\sim \equiv(\mu A)^{\dg\T} =\mu^* b^\dg\times a^\dg$. When it acts on the product of superoperators, we have $(AB)^\sim=\tilde{A}\tilde{B}$, note that the order of the superoperators is not changed.

These three operations are the inverse of themselves.

\vspace{6pt}

We can now write down the required conditions on general transformation \cite{Sudarshan61,Weinberg14}. We consider a general transformation $S$ with the exponential form
%%%
\begin{align}   \label{SG}
    S\equiv e^{-\th G}\,,
\end{align}
%%%
where $G$ is its generator and $\th$ is a real parameter. Its inverse is $S^{-1}=e^{\th G}$. Under the general transformation, density matrices transform as
%%%
\begin{align}   \label{rho'}
\rho'\equiv S\rho\,.
\end{align}
%%%
To preserve the hermiticity of density matrices, general transformation should be adjoint-symmetric, defined by $\tilde{S}=S$. This can be shown by using the operations introduced at the beginning of this section. We find that
%%%
\begin{align}   \label{adjsym}
    (\rho')^\dg=(S\rho)^\dg=\rho^\dg S^\dg=(S^\dg)^\T\rho=\tilde{S}\rho\,.
\end{align}
%%%
Setting this equal to \Eq{rho'}, we obtain the desired result. In terms of the generators, this is equivalent to
%%%
\begin{align}   \label{Gadjsym}
        \tilde{G}&=G\,.&\quad &\text{(hermitian condition)}
\end{align}
%%%

The trace of density matrices is preserved whenever $S$ satisfies $\tr(S\rho)=\tr(\rho')=1$. This translates into a condition on the generator as
%%%
\begin{align}   \label{Gtrace}
        \tr(G\rho)&=0 &\quad &\text{(trace condition)}
\end{align}
%%%
for arbitrary $\rho$, which can be shown by expanding the exponential in polynomial form.

The positivity of density matrices requires $\rho'>0$. This condition cannot be simplified further in terms of the generator. We know that completely positive maps satisfy this requirement. For not completely positive maps, we can use this condition to determine the valid range of the transformation parameter. The last three conditions on the general transformation are similar to those given by Ref.~\cite{Sudarshan61,Weinberg14} in component form for dynamical maps. In Sec.~\ref{Sec2Lev}, we find that the hermiticity and trace condition are sufficient to determine the form of the generators, whereas the positivity condition limit the range of the parameter of transformation. The same conclusion applies to the general transformation in continuous system \cite{Tay17}.

The generators that satisfy the hermitian and trace condition are closed under the commutator bracket. This can be proved by showing that the commutator bracket between two generators, $F$ and $G$, that satisfy the hermitian and trace condition, is adjoint-symmetric,
%%%
\begin{align}   \label{[AB]}
        [F,G]^{\widetilde{}}=\tilde{F}\tilde{G}-\tilde{G}\tilde{F}=FG-GF=[F,G]\,,
\end{align}
%%%
and satisfies the trace condition,
%%%
\begin{align}   \label{trFG}
        \tr([F,G]\rho)=\tr\big(F (G\rho)\big)-\tr\big(G (F\rho)\big)=0\,.
\end{align}
%%%
The commutator $[F,G]$ therefore belongs to the same space of the generators and can be decomposed into a linear sum of them. This allows us to apply the theory of Lie group \cite{Gilmore} to general transformation as usually done in ordinary symmetry.

While density matrices transform as in \Eq{rho'}, the invariance of the expectation value requires the observable to transform as
%%%
\begin{align}   \label{a'}
        a'= S^{-1}_\adj a\,,
\end{align}
%%%
where
%%%
\begin{align} \label{Sadj}
    S_\adj\equiv S^\T\,.
\end{align}
%%%
This can be seen from the invariance of the expectation value by moving the inverse of $S$ to the left of the observables
%%%
\begin{align}   \label{expect'}
    \<a\>_\rho&=\tr\big(a\cdot S^{-1}S\rho\big)=\tr\big((S^{-1})^\T a \cdot \rho'\big)=\< a'\>_{\rho'} \,.
\end{align}
%%%
This is the same as the definition of the adjoint operator $\hat{K}$ and $\mathcal{L}^\dg$ defined in Refs.~\cite{Fujiwara99} and \cite{Breuer}, respectively.
Note that $S_\adj$ is adjoint-symmetric, $\tilde{S}_\adj=S_\adj$.
There is a requirement on the generator obtained from a constraint on $S^{-1}_\adj$. Consider the situation when $a$ is the $N\times N$ identity operator $1_N$, $\<1_N\>_\rho=\<1_N'\>_{\rho '}=1$ would require $1_N'=S^{-1}_\adj 1_N = 1_N$ for consistency. This gives $G^\T 1_N=0$.

We will now derive the condition under which a generator will give rise to compact transformation. We know that compact transformation permits unitary representation, $S^\dg=S^{-1}$. Hence, its generator is anti-hermitian $G^\dg=-G$. The hermitian condition \eqref{Gadjsym} is equivalent to $G^\T=\tilde{G}^\T=(G^{\dg\T})^\T=G^\dg$, since transposition is its own inverse. Putting the results together, we find that the generator is required to be both anti-hermitian and anti-symmetric
%%%
\begin{align}   \label{Gunitary}
    G^\dg=G^\T=-G\,. &\quad &\text{(unitary condition)}
\end{align}
%%%

%%%%%%%%%%%%%%%%%%%%%%%%%%
%       Section          %
%%%%%%%%%%%%%%%%%%%%%%%%%%
\section{General transformation in two-level system}
\label{Sec2Lev}

In two-level system, we parameterize density matrix as
%%%
\begin{align}   \label{rho2}
    \rho=\frac{1}{2}(1_2+\vec{r}\cdot \vec{\s})=
    \frac{1}{2} \left(\begin{array}{cc}
            1+z & x-\i y \\
            x+\i y & 1-z
          \end{array}
    \right)\,,
\end{align}
%%%
where $ \vec{\s}=(\s_1,\s_2,\s_3)$ in which $\s_i$ are the Pauli matrices, and $\vec{r}\equiv (x,y,z)$ is the Bloch vector. The coordinates of the Bloch vector are defined as usual $x=\<\s_1\>_\rho, y=\<\s_2\>_\rho$ and $z=\<\s_3\>_\rho$.
The density matrix is positive if the Bloch vector lies in the Bloch ball \cite{Nielsen}, $
\vec{r}\,^2=x^2+y^2+z^2\leq 1$.

Generators that satisfy the hermitian \eqref{Gadjsym} and trace condition \eqref{Gtrace} can be constructed from the Pauli matrices as
%%%
\begin{subequations}
\begin{align}   \label{Ri2}
    \i R_i&\equiv \frac{\i}{2} (\s_i\times 1_2-1_2\times \s_i)\,,\\
    D_i &\equiv \frac{1}{2}(\s_i\times \s_i-I_2)\,,\label{Hii2}\\
    H_{ij}&\equiv \frac{1}{2}(\s_i\times \s_j+\s_j\times \s_i) \,,\qquad i\neq j\,,\label{Hij2}\\
    P_{ij}&\equiv \frac{\i}{2} (\s_i\times \s_j-\s_j\times \s_i)-\frac{1}{2}\epsilon_{ijk}(\s_k\times 1_2 +1_2\times \s_k)\,,    \label{Pij2}
\end{align}
\end{subequations}
%%%
where $i, j =1,2,3$, $I_2\equiv  1_2\times 1_2$, in which $1_2$ is the $2\times 2$ identity matrix, and $\epsilon_{ijk}$ is the total antisymmetric tensor with $\epsilon_{123}=1$. To prove that the generators \eqref{Ri2}-\eqref{Pij2} indeed satisfy the hermitian and trace condition, we need to make use of the definition of the association operation, as well as the invariance of trace under cyclic permutations of operators and simplify the products of Pauli matrices through the identity $\s_i\s_j=\del_{ij}+\i \eps_{ijk}\s_k$. We note that $H_{ij}$ and $P_{ij}$ are symmetric and anti-symmetric under the permutation of $i, j$, respectively.

The general transformation in two-level system should send a density matrix to another in the Bloch ball to satisfy the positivity requirement. This constraint gives rise to state-dependent condition on the parameter of transformation.

The generators \eqref{Ri2}-\eqref{Pij2} give rise to four types of general transformation based on their actions on the Bloch vector. Generic transformations can be decomposed into them.

\vspace{6pt}

\nd \textit{(1) Rotation.}
%\label{SecRot}
The generator $\i R_i$ gives rise to ordinary symmetry $e^{-\th \i R_i}=e^{-\i \th \s_i/2}\times e^{\i \th \s_i/2}$. Using $e^{-\i\th \s_i/2}=\cos(\th/2)1_2-\i\sin(\th/2)\s_i$, we can write it as a sum of the generators
%%%
\begin{align}   \label{expR}
    e^{-\th \i R_i}&= I_2-\sin\th \,\i R_i+(1-\cos\th )D_i \,.
\end{align}
%%%
It rotates the Bloch vector along the $i$-axis counter-clockwise for positive $\th$, while keeping its length unchanged.
The positivity requirement is always fulfilled, leaving no constraint on the parameter.
As an example, we consider $i=3$. The transformed density matrix $\rho'=e^{-\th \i R_3}\rho$ has the Bloch vector
%%%
\begin{align}   \label{Rrho}
        \vec{r}\, '=(x\cos\th-y\sin\th, x\sin\th +y\cos \th, z)\,.
\end{align}

\nd \textit{(2) Dilation.}
%\label{SecDilation}
By simplifying the products of Pauli matrices, we find that $(-D_i )^n=-D_i , n=1,2,\cdots$. Consequently,
%%%
\begin{align}   \label{expHii}
    e^{-\mu D_i }&=\sum_{n=0}^\infty \frac{\mu^n}{n!}(-D_i )^n
    =I_2+\left(1-e^{\mu}\right)D_i \,.
\end{align}
%%%
This equation reveals two features of the general transformation to us. Firstly, general transformation is in general not factorized, i.e., it cannot be cast into the form $u\times u^\dg$ of ordinary symmetry, where $u$ is a unitary transformation. Under ordinary symmetry, pure states remain pure. On the other hand, general symmetry sends pure states into mixed states. In other words, general transformation is not distributive. Non-distributiveness is also a property of the so-called star-unitary transformation which produces irreversibility in Poincar\'e nonintegrable systems from the full dynamics \cite{Prigogine73,Kim03}.

Secondly, general transformation in general is not completely positive as exemplified by the existence of negative coefficient of $D_i $ for $\mu>0$ \cite{Sudarshan03}.
A positive $\mu$ results in a dilation of the Bloch vector in the plane perpendicular to the $i$-axis away from the axis. There is danger in dilating the Bloch vector outside the Bloch ball unless $\mu$ satisfies the inequality
%%%
\begin{align}   \label{Blochposr'}
    \vec{r}\,'^2=x'^2+y'^2+z'^2\leq 1\,,
\end{align}
%%%
where the transformed coordinates are defined by $x'\equiv \<\s_1\>_{\rho'}$, and etc.
For example, if $i=3$, $\rho'=e^{-\mu D_3 }\rho$ gives
%%%
\begin{align}   \label{H33rho}
    \vec{r}\, '=(e^{\mu}x\,, e^{\mu}y\,, z)\,.
\end{align}
%%%
For $\mu<0$, it is a contraction and all points on the 3-axis are fixed points of transformation.

\vspace{6pt}

\nd \textit{(3) Hyperbolic rotation.}
%\label{SecHrot}
We simplify the products of Pauli matrices to obtain $H_{ij}^2=-D_k $ and $H_{ij}(-D_k )=H_{ij}$, where $k$ satisfies $|\eps_{ijk}|=1$. This generalizes to
$H_{ij}^{2n}=-D_k $ and $H_{ij}^{2n+1}=H_{ij}$, for positive integer $n$. Then,
%%%
\begin{align}   \label{expHij}
    e^{-\phi H_{ij}}&=I_2+\sum_{n=1}^\infty \frac{(-\phi)^{2n}}{(2n)!}H_{ij}^{2n}
        +\sum_{n=0}^\infty \frac{(-\phi)^{2n+1}}{(2n+1)!}H_{ij}^{2n+1}\nm
    &=I_2+\left(1-\cosh\phi\right)D_k -\sinh \phi \,H_{ij}\,.
\end{align}
%%%
It causes the Bloch vector to undergo a hyperbolic rotation in the $ij$-plane. For example, $ \rho'=e^{-\phi H_{12}}\rho$ gives
%%%
\begin{align}   \label{H12rho}
    \vec{r}\, '=(x\cosh\phi -y\sinh\phi \,, -x\sinh\phi +y\cosh\phi \,,z)\,.
\end{align}
%%%
The inequality \eqref{Blochposr'} determine the valid range of $\phi$.

\vspace{6pt}

\nd \textit{(4) Translation.}
%\label{SecTrans}
We find that $P_{ij}^2=0$. As a result,
%%%
\begin{align}   \label{expPij}
    e^{-\zeta P_{ij}}&=I_2- \zeta P_{ij}\,.
\end{align}
%%%
It translates the Bloch vector in the direction perpendicular to the $ij$-plane. For example, $\rho'=e^{-\zeta P_{12}}\rho$ gives
%%%
\begin{align}   \label{P12rho}
    \vec{r}\, '=(x\,, y\,, z+\zeta)\,.
\end{align}
%%%
The range of $\zeta$ should be restricted to satisfy \Eq{Blochposr'}.

\vspace{6pt}

Testing the generators \eqref{Ri2}-\eqref{Pij2} for the unitary condition \eqref{Gunitary}, we find that ordinary symmetry (rotation) is the only compact transformation in two-level system. The rests of the general transformation do not have factorized form and are not compact. The same conclusion is reached when the generators are generalized to $N$-level systems in Sec.~\ref{SecNLevel}.

The corresponding adjoint transformation on the observables, such as $\s'_i=S_\adj \s_i$, can be inferred from each of the transformed Bloch vector $\vec{r}\,'$ by using the invariance of expectation value under the general transformation. It can also be obtained by calculating $S_\adj\s_i$ directly.

Let us discuss the results in this section in reference to previous works. In Ref.~\cite{Fujiwara99}, it was shown that the dimensions of completely positive quantum operations are 12 for two-level system. This is consistent with the number of generators of the general transformation. While the time evolution operator for open systems is irreversible and hence it forms semigroup, the general transformation is kinematical, i.e., it does not involve the dynamics. Therefore, it can transform the density matrix in the forward and backward direction with respect to the parameter.

We can use the Fujiwara-Algoet condition \cite{Fujiwara99} to test the completely positive nature of the general transformation. We first write the transformed Bloch vector as $\vec{r}\,'=\text{A}\vec{r}+\vec{\kappa}$ \cite{Fujiwara99,Oi01,Zyczkowski04}. $\text{A}$ is a $3\times 3$ real matrix and its singular values are denoted by the components of the vector $\vec{\eta}=(\eta_x, \eta_y, \eta_z)$. For unital maps, $\vec{\kappa}=\vec{0}$, the transformation is completely positive if and only if the singular values satisfy the Fujiwara-Algoet conditions $(\eta_x\pm \eta_y)^2\leq (1\pm \eta_z)^2$ \cite{Fujiwara99,Oi01}. The ordinary rotation with $\vec{\eta}=(1,1,1)$ is obviously completely positive. The dilation transformation with $\vec{\eta}=(e^\mu, e^\mu, 1)$ will be completely positive provided $\mu\leq 0$. As for the hyperbolic transformation, $\vec{\eta}=(e^\phi, e^{-\phi}, 1)$ implies that the Fujiwara-Algoet condition cannot be satisfied except trivially. Hence, the hyperbolic transformation cannot be completely positive. The translation can bring the Bloch vector outside the Bloch ball. Hence, it is also not completely positive.

There had been debates on whether quantum operations must be completely positive \cite{Pechukas94,Shaji05}. It was understood that not completely positive maps would appear for systems initially entangled to the environment \cite{Shaji04}. Therefore, positive but not completely positive maps could also be physically relevant. Methods were devised to remove negativity through the slippage of initial conditions, or one could limit the class of density matrices on which not completely positive maps could act on to ensure the positivity of density matrices \cite{Shaji04,Shaji05}.

The general transformation we construct can rotate the Bloch vector either circularly or hyperbolically, dilate it or translate it. On the other hand, the completely positive quantum operations \cite{Ruskai01,Zyczkowski04} give different geometrical pictures when we consider their actions on the Bloch ball. They map the Bloch ball into an ellipsoid, where the parameters in the quantum operations determine the size of the ellipsoid, the orientation of its axis, and the position of its center.

%%%%%%%%%%%%%%%%%%%%%%%%%%i
%       Section          %
%%%%%%%%%%%%%%%%%%%%%%%%%%
\section{Symmetry of reduced dynamics}
\label{SecInv}

In this section, we consider the symmetry of the generator of reduced dynamics. In particular, we consider the symmetry of amplitude damping and phase damping \cite{Nielsen} for two-level system under the general transformation. Under a similarity transformation by $S$, the equation of motion $\d\rho/\d t=-K\rho$ goes into $\d\rho'/\d t=-K'\rho'$, where
%%%
\begin{align}   \label{simtransf}
        K'&\equiv S K S^{-1}\,.
\end{align}
%%%
If $K'$ has the same form as $K$, then $S$ is a symmetry of $K$ and $\rho'$ is also a solution to $K$.

%%%%%%%%%%%%%%%%%%%%%%%%%%
%       Section          %
%%%%%%%%%%%%%%%%%%%%%%%%%%
\subsection{Amplitude damping}
\label{SecAmpDamp}

Amplitude damping describes the relaxation of a two-level system \cite{Nielsen}, for example, a laser system in contact with a thermal bath. It is initiated by the generator
%%%
\begin{align}   \label{ampdamp}
     K_\amp &\equiv\i\frac{\w_0}{2}[\s_3,\rho]
    -\frac{\gam}{2}n(2\s_+\rho\s_--\s_-\s_+\rho-\rho\s_-\s_+)\nm
    &\quad
    -\frac{\gam}{2}(n+1)(2\s_-\rho\s_+-\s_+\s_-\rho-\rho\s_+\s_-)\,,
\end{align}
%%%
in which $\w_0$ is the natural frequency of the system, $\gam$ is the relaxation rate, $n$ is the Bose-Einstein distribution $n\equiv 1/\big(e^{\w_0/k_\text{B}T}-1\big)$, where $T$ is the temperature of the bath, and $\s_\pm\equiv \frac{1}{2}(\s_1\pm \i \s_2)$.
In terms of the set of generators introduced in the last section, $K_\amp$ is cast into the form
%%%
\begin{align}   \label{ampdampRPH}
    K_\amp(b,\gam) &=K_0+K_d(b,\gam)\,,
\end{align}
%%%
where
%%%
\begin{align} \label{K0}
    K_0 &\equiv\w_0\i R_3\,,
\end{align}
%%%
%%%
\begin{align} \label{Kd}
    K_d(b,\gam) &\equiv -\gam b \left(\frac{1}{2b}P_{12}+D_1 +D_2 \right)\,,
\end{align}
%%%
%%%
\begin{align}
    b&\equiv n+\frac{1}{2}=\frac{1}{2}\coth\left(\frac{\w_0}{2k_\text{B}T}\right)\,.
\end{align}
%%%

%%%%%%%%%%%%%%%%%%%%%%%%%%
%       Section          %
%%%%%%%%%%%%%%%%%%%%%%%%%%
\subsection{Solution to amplitude damping}
\label{SecSolAmp}

Let us first present the solution to amplitude damping.
In the interaction picture defined by $\bar{\rho}\equiv e^{\w_0 t\i R_3}\rho$, the equation of motion is $\d\bar{\rho}/\d t=-\bar{K}_\amp \bar{\rho}$, with the generator
%%%
\begin{align}   \label{Kampint}
        \bar{K}_\amp&\equiv e^{\w_0 t\i R_3}K_d e^{-\w_0 t\i R_3}\nm
                &=K_d+\w_0t[\i R_3,K_d]+\frac{(\w_0 t)^2}{2!}[\i R_3, [\i R_3,K_d]]+\cdots\nm
                &=K_d\,.
\end{align}
%%%
since $\i R_3$ commutes with $P_{12}$ as well as $D_1 +D_2 $.
The fact that $\frac{1}{4b}P_{12}+D_1 $ commutes with $\frac{1}{4b}P_{12}+D_2 $ then allows one to separate the dissipative component into two parts,
%%%
\begin{align}   \label{Kampintexp}
e^{-\bar{K}_\amp t}
=e^{\gam b t (\frac{1}{4b}P_{12}+ D_2 )}e^{\gam b t (\frac{1}{4b}P_{12}+ D_1 )}\,.
\end{align}
%%%
Using the relation $\big(\frac{1}{4b}P_{12}+ D_i \big)^2=-\big(\frac{1}{4b}P_{12}+D_i \big)$ for $i=1,2$, we deduce that $\big(\frac{1}{4b}P_{12}+ D_i \big)^n=(-1)^{n-1}\big(\frac{1}{4b}P_{12}+D_i \big)$ for $n=1, 2, ,3, \cdots$. We then obtain
%%%
\begin{align}   \label{eP12}
        e^{\gam b t (\frac{1}{4b}P_{12}+ D_i )}
        &=I_2+\sum_{n=1}^\infty \frac{(\gam b t)^n}{n!} \left(\frac{1}{4b}P_{12}+ D_i \right)^n \nm
        &=I_2+\left(1-e^{-\gam b t}\right)\left(\frac{1}{4b}P_{12}+D_i \right)\,.
\end{align}
%%%
Multiplying the exponentials in \Eq{Kampintexp} then yields
%%%
\begin{align}   \label{eP12eP12}
        e^{-\bar{K}_\amp t}
%            &=e^{\frac{1}{2}\gam t (\frac{1}{2}P_{12}+b D_2 )}e^{\frac{1}{2}\gam t (\frac{1}{2}P_{12}+b D_1 )}\nm
            &=I_2+\frac{1}{2}\left(1-e^{-2\gam b t}\right)\left(\frac{1}{2b}P_{12}+D_1 +D_2 \right)\nm
            &\quad+\frac{1}{2}\left(1-e^{-\gam b t}\right)^2D_3 \,,
\end{align}
%%%
where we make use of $D_1 P_{12}=D_2 P_{12}=-P_{12}$, $P_{12}D_1 =P_{12}D_2 =0$ and $D_1 D_2 =D_2 D_1 =\frac{1}{2}(D_3 -D_1 -D_2 )$.
\Eq{eP12eP12} explicitly shows that the time evolution operator can be decomposed into a linear sum of the hermitian and trace preserving generators with positive coefficients for $t>0$.

As a consequence of \Eq{eP12eP12}, $\bar{\rho}(t)=e^{-\bar{K}_\amp t}\bar{\rho}_0$ gives the following time evolution of the Bloch vector in the interaction picture
%%%
\begin{align}   \label{ampBloch}
    \vec{\bar{r}}=\big(x_0 e^{-\gam bt}\,, y_0 e^{-\gam bt}\,, z_0e^{-2\gam bt}-\frac{1}{2b}(1-e^{-2\gam bt}) \big)\,.
\end{align}
%%%
The trajectory traced out by the Bloch vector during its time evolution in the interaction picture is illustrated by the black curve in \Fig{fig1}(c).
The time evolution in the Schr\"odinger picture is obtained through a rotation in the 3-axis via an angle $\w_0t$, cf.~\Eq{Rrho}, as illustrated by the black curves in \Figs{fig1}(a), \ref{fig1}(b) and \ref{fig1}(d). In the limit $t\rightarrow \infty$, the stationary state in both pictures is the Gibbs state with $x_\st=0, y_\st=0, z_\st=-1/(2b)$.

%%%%%%%%%%%%%%%%%%%%%%%%%%
%       Section          %
%%%%%%%%%%%%%%%%%%%%%%%%%%
\subsection{Exact and form invariant symmetry}
\label{SecSymAmp}

We distinguish two types of symmetry, (1) exact symmetry when $K'=K$, and (2) form invariant symmetry when $K'$ has the same form as $K$ but with different coefficients.
Since $\i R_3$ and $D_3 $ commute with $K_\amp$, $e^{-\th\i R_3}$ and $e^{-\mu D_3 }$ are exact symmetry of $K_\amp$, $K'_\amp=K_\amp$. We can show this by a straight-forward calculation,
%%%
\begin{align}
    &K'_\amp=e^{-\th\i R_3}K_\amp e^{\th\i R_3}\nm
        &\quad=K_\amp-\th[\i R_3,K_\amp]+\frac{\th^2}{2!}[\i R_3,[\i R_3,K_\amp]]+\cdots\nm
        &\quad=K_\amp\,,
\end{align}
%%%
and similarly for the general transformation under $e^{-\mu D_3 }$.

\begin{figure}[t]
  \centering
\includegraphics[width=7.6cm,trim={5.4cm 9cm 5.6cm 10cm},clip]{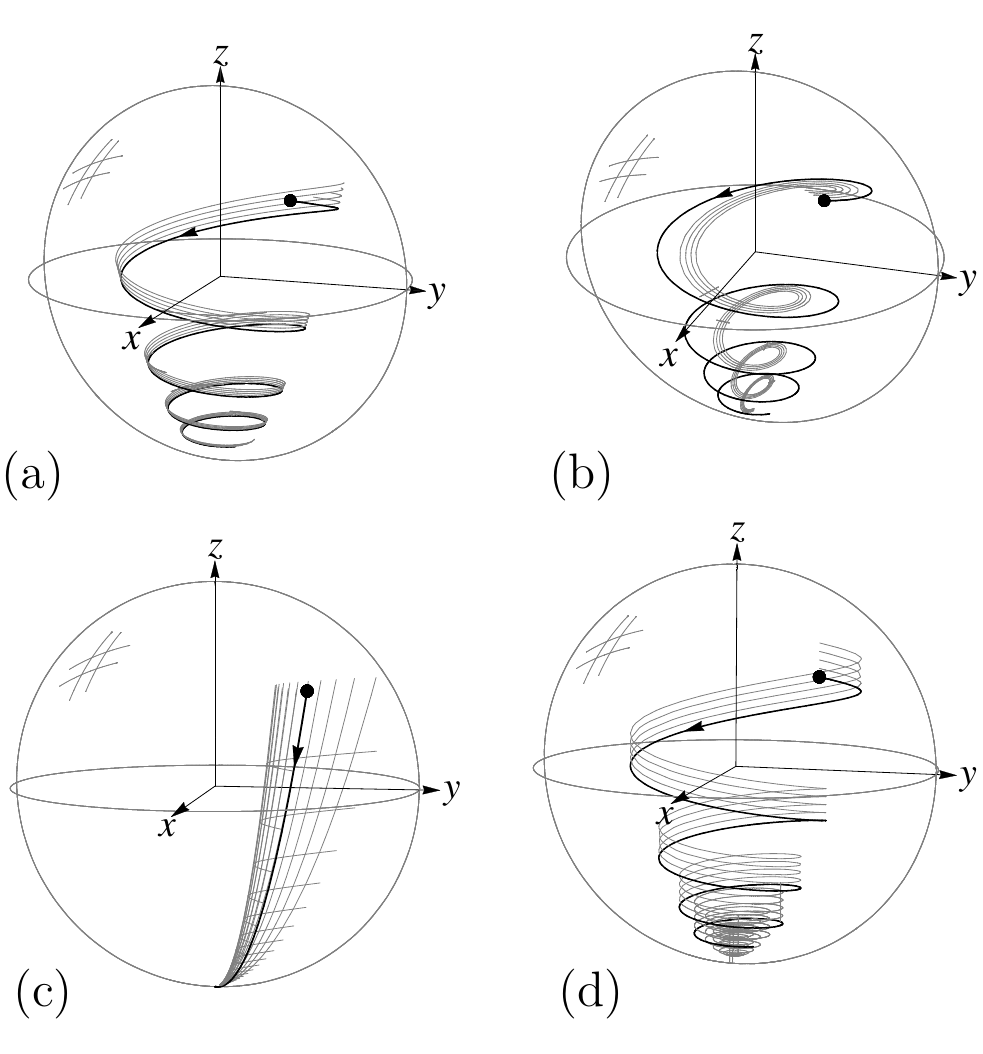}
\caption{Amplitude damping under general transformation. The initial condition is $(x_0, y_0, z_0)=(0.4,0.5,0.5)$, and we use the parameters $\w_0=1$, $b=0.5$ and $\gam=0.1$. There are two types of symmetry. (1) Exact symmetry: (a) rotation by $\exp(-\th \i R_3)$ and (b) dilation by $\exp(-\mu D_3 )$ for $\mu<0$ in the Schr\"odinger picture, and (c) hyperbolic rotation by $\exp(-\phi H_{12})$ in the interaction picture. (2) Form invariant symmetry: (d) translation by $\exp(-\zeta P_{12})$ in the Schr\"odinger picture. Black curves in (a)-(d) refer to the same specific solution of amplitude damping in the Schr\"odinger picture (a), (b), (d), and in the interaction picture (c). For exact symmetry, gray curves show a family of solutions to $K_\amp$ in (a), (b), and to $\bar{K}_\amp$ in (c), obtained by acting the transformation on the specific solution. For form invariant symmetry (d), gray curves are solutions to $K(b',\gam')$ that correspond to different $\zeta$, cf.~\Eqs{zetab}-\eqref{zetagam}, each ends up in different stationary states.}
\label{fig1}
\end{figure}

This implies that $\rho'(t)$ forms a family of solutions to the equation of motion, with the Bloch vector transforms according to \Eqs{Rrho} and \eqref{H33rho}, respectively. Under $e^{-\th\i R_3}$, the trajectory of its evolution with time is rotated by an angle $\th$ along the 3-axis, whereas under $e^{-\mu D_3 }$ for $\mu<0$, the trajectory is contracted in the 12-plane towards the 3-axis, for illustrations see the families of gray curves in \Fig{fig1}(a) and \ref{fig1}(b), respectively.

The generator of hyperbolic rotation $H_{12}$ does not commute with the unitary component of $K_\amp$. However, we find that it commutes with the dissipative component, $[H_{12}, \bar{K}_d]=0$, by using $[H_{12},P_{12}]=0=[H_{12},D_1+D_2]$. Consequently, by \Eq{Kampint}, $e^{-\phi H_{12}}$ furnishes an exact symmetry of amplitude damping only in the interaction picture, $\bar{K}'_\amp=\bar{K}_\amp$.
The trajectory of the time evolution of the Bloch vector in the interaction picture is transported along section of a hyperbolic curve by an angle $\phi$, as illustrated by the family of gray curves in \Fig{fig1}(c).

A translation of $K_\amp$ by $e^{-\zeta P_{12}}$ yields a form invariant symmetry. Using $[P_{12},\i R_3]=0$ and $[P_{12},D_1 +D_2 ]=2 P_{12}$, we find that $[P_{12},K_\amp]=-2\gam b  P_{12}$. Hence,
%%%
\begin{align}
   &K'_\amp=e^{-\zeta P_{12}}K_\amp e^{\zeta P_{12}}\nm
        &\quad=K_\amp -\zeta [P_{12},K_\amp]+\frac{\zeta^2}{2!}[P_{12},[P_{12},K_\amp]]+\cdots\nm
        &\quad=K_\amp+2\gam b \zeta P_{12}\nm
        &\quad=\w_0\i R_3-\gam b
        \left( \frac{1-4b\zeta}{2b}P_{12}+D_1 +D_2\right)\,.
\end{align}
%%%
It is now possible to cast $K'_\amp$ in the same form as the original dynamics $K_\amp$ but with different coefficients, so that this furnishes a form invariant symmetry. By setting $K'_\amp=K_\amp(b',\gam')$, we obtain two conditions on the temperature $b'$ and relaxation rate $\gam'$ of the transformed system. One of them allows us to set the new temperature parameter as
%%%
\begin{align} \label{zetab}
    b'=\frac{b}{1-4b\zeta}\,,
\end{align}
%%%
whereas the other one can be satisfied provided we choose the relaxation rate as
%%%
\begin{align} \label{zetagam}
    \gam'&=\frac{b}{b'}\gam=(1-4b\zeta)\gam \,.
\end{align}
%%%

%%%
%%%

The transformed system now evolves as in \Eq{eP12eP12} with different temperature $b'$ and different relaxation rate $\gam'$, though the effective relaxation rate $\gam b=\gam'b'$ remains constant. This is the two-level system generalization of the thermal symmetry discussed in the harmonic oscillator system \cite{Tay17}. The trajectory of the time evolution of the Bloch vector is translated by a distance of $\zeta$ along the positive 3-axis, as illustrated by the family of gray curves in \Fig{fig1}(d). We note that the trajectories with different $\zeta$ end up in different stationary states.

Ref.~\cite{Zyczkowski04} discussed a different symmetry initiated by two ordinary transformations in the form (in our notations) $K\rightarrow K_{vw}\equiv (v\times v^\dg)K(w\times w^\dg)$, where $v$ and $w$ are unitray operators. $K$ and $K_{vw}$ are in general different quantum operations, i.e., their form change under the symmetry. However, they are called unitarily similar because the spectra of their dynamical matrices remain invariant. Another closely related work in Ref.~\cite{Sudarshan06} identified the group structure of dynamical maps to be $u\Delta v^\dg$, which acts on the Bloch vector of $N$-level systems, where $u, v$ are unitary transformations from $SU(N)$, and $\Delta$ is a classical stochastic semigroup. In contrast, the symmetry we consider is a similarity transformation \eqref{simtransf} that keeps the forms of the quantum operations invariant.

%%%%%%%%%%%%%%%%%%%%%%%%%%
%       Section          %
%%%%%%%%%%%%%%%%%%%%%%%%%%
\subsection{Phase damping}
\label{SecPhDamp}

Phase damping in two-level system is generated by
%%%
\begin{align}   \label{phdamp}
    K_\ph\equiv-\frac{\gam}{2}(\s_3\times\s_3-I_2) \,.
\end{align}
%%%
In terms of the generators we introduced, it takes a simple form
%%%
\begin{align}   \label{phdampGen}
    K_\ph=-\gam D_3 \,.
\end{align}
%%%
Phase damping causes the coherence of the system to decay over time. From \Eq{expHii}, the time evolution operator is
%%%
\begin{align}
    e^{-K_\ph t}=e^{\gam t D_3 }&=I_2+(1-e^{-\gam t})D_3 \,.
\end{align}
%%%
The solution to the equation of motion has a simple form,
%%%
\begin{align}   \label{phsol}
    \vec{r}\,'=(x_0 e^{-\gam t}\,, y_0 e^{-\gam t}\,, z_0)\,.
\end{align}
%%%
Since all of $\i R_3, D_3 , H_{12}$ and $P_{12}$ commute with $K_\ph$, $K'_\ph=K_\ph$, the transformations they generate are exact symmetries of phase damping.

%%%%%%%%%%%%%%%%%%%%%%%%%%
%       Section          %
%%%%%%%%%%%%%%%%%%%%%%%%%%
\subsection{Stationary state in two-level system}
\label{SecStN2}

Now we consider the generic generator of the reduced dynamics of two-level system and obtain its stationary state $\rho_\st$. In particular, we consider $\rho_\st$ with zero off-diagonal components, i.e., $x_\st=y_\st=0$. The Gibbs state is an example of states with zero coherence.
The most general generator of time evolution is a linear sum of the generators over real coefficients
%%%
\begin{align}   \label{genK}
    K_\text{gen}&=\w_0 \,\i R_3+\sum_{i=1}^{3}\al_{ii}D_i+\sum_{i< j=1}^{3}\big(\al_{ij}H_{ij}+\bt_{ij}P_{ij}\big)\,,
\end{align}
%%%
where we assume that the unitary part is already diagonalized.
Stationary state satisfies $K\rho_\st=0$, from which we obtain
%%%
\begin{align}   \label{nst}
    z_\st=-\frac{2\bt_{12}}{\al_{11}+\al_{22}}=-\frac{2\bt_{13}}{\al_{23}}
    =\frac{2\bt_{23}}{\al_{13}}\,.
\end{align}
%%%
Hence, the stationary state is independent of $\w_0, \al_{12}$, and $\al_{33}$. From \Eq{nst}, we learn that if $\al_{11}+\al_{22}, \al_{23}$, or $\al_{13}$ vanish separately, the coefficients of the $P_{12}, P_{13}$, or $P_{23}$ terms have to vanish separately from the reduced dynamics as well, respectively. This is indeed true in amplitude damping and phase damping.

%%%%%%%%%%%%%%%%%%%%%%%%%%
%       Section          %
%%%%%%%%%%%%%%%%%%%%%%%%%%
\section{Generalization to $N$-level system}
\label{SecNLevel}

The set of generators that produces hermiticity and trace-perserving general transformation can be generalized to $N$-level system through the generalized $\lam$-matrices \cite{Tilma02}, $\lam_i, i=1,2,\cdots,N^2-1$, for $N\geq 3$. The $\lam$-matrices have the product
%%%
\begin{align}   \label{tt}
    \lam_i\lam_j=\frac{1}{2N}\del_{ij}1_N+\frac{1}{2}\sum_{k=1}^{N^2-1} d_{ijk}\lam_k+\frac{\i}{2}\sum_{k=1}^{N^2-1} f_{ijk}\lam_k\,,
\end{align}
%%%
where $d_{ijk}$ and $f_{ijk}$ are totally symmetric and totally anti-symmetric tensors in all pairs of indices, respectively. The matrices have the commutation relation $[\lam_i,\lam_j]=\i f_{ijk}\lam_k$.
The $f$- and $d$-tensors can be obtained from
%%%
\begin{align}   \label{f}
    f_{ijk}&=-2\i \,\tr(\lam_k[\lam_i,\lam_j])\,,\\
    d_{ijk}&=2 \,\tr(\lam_k\{\lam_i,\lam_j\})\,, \label{d}
\end{align}
%%%
where curly bracket denotes anti-commutator.
The $d$-tensor vanishes in two-level system.

The generators that obey the hermitian \eqref{Gadjsym} and trace \eqref{Gtrace} condition are
%%%
\begin{subequations}
\begin{align}   \label{Ri}
    \i R_i&\equiv \i(\lam_i\times 1_N-1_N\times \lam_i)\,,\\
    H_{ij}&\equiv 2T_{\!\!ij}-d_{ijk}L_k-\frac{2}{N}\del_{ij}I_N\,,\label{Hij}\\
    P_{ij}&\equiv 2\i A_{ij}-f_{ijk}L_k\,,    \label{Pij}
\end{align}
\end{subequations}
%%%
where
%%%
\begin{subequations}
\begin{align}
    I_N&\equiv 1_N\times 1_N\,,\\
    L_k&\equiv \lam_k\times 1_N+1_N\times \lam_k\,,\\
    T_{ij}&\equiv \lam_i\times \lam_j+\lam_j\times \lam_i\,,\label{S}\\
    A_{ij}&\equiv \lam_i\times \lam_j-\lam_j\times \lam_i\,.
\end{align}
\end{subequations}
%%%
Note that the dilation $D_i$ discussed in the two-level system is hidden as $H_{ii}$ in its generalization to \Eq{Hij}. The fact that these generators satisfy the hermitian and trace conidition can be proved straight-forwardly using the invariance of the cyclic permutations of matrices under the trace and \Eq{tt}. From \Eqs{Ri}-\eqref{Pij}, we verify that only $\i R_i$ satisfies the unitary condition \eqref{Gunitary}. Of all the general transformations, only the ordinary unitary rotations are factorized and compact.

There are $N^4-N^2$ generators, consistent with the dimensions of the convex set of completely positive quantum operations \cite{Fujiwara99,Zyczkowski04}. They give a complete list of the generators that yield all the hermiticity and trace preserving general transformations in the $N$-level system. The generator of reduced dynamics is a linear sum of them over real coefficients, see App. \ref{SecGenNLev} for details. We also list the commutation relations between the generators in the appendix.

%%%%%%%%%%%%%%%%%%%%%%%%%%
%       Section          %
%%%%%%%%%%%%%%%%%%%%%%%%%%
\section{Conclusions}
\label{SecConclusion}

We have studied the kinematical symmetry of reduced dynamics in the density matrix space. For this purpose, we construct a complete set of generators which produces hermiticity and trace preserving general transformation on density matrix. Together with the corresponding adjoint transformation on the observable, the general transformation keep the expectation value of observable invariant. We need to emphasize that these are generators in the superoperator space. Except for the ordinary symmetry (rotation) generated by unitary operators, the general transformation are neither factorized nor compact, and in general are not completely positive. However, they can be made positive by proper choice of the transformation parameters. We want to emphasize that positive but not completely positive maps are also physically relevant as has been debated in the literature.

By writing the generators in terms of superoperators, we can directly apply the method of group theory to general symmetry in the density matrix space in a similar way to the applications of group theory to ordinary symmetry in the state vector space. Consequently, we are able to reduce the non-factorized transformation to a simple form. The generators can also be generalized to arbitrary higher level systems. The general transformations in the two-level system result in simple actions on the Bloch vector, namely ordinary and hyperbolic rotations, dilations and translations. With respect to the symmetry of the reduced dynamics in two-level system, we discuss two types of symmetry, i.e., exact symmetry and form invariant symmetry. It is interesting to consider the general symmetry in multi-particle systems and explore their applications in the future.

In memory of Professor E. C. G. Sudarshan, whose works have always been sources of inspiration to us.

\appendix

%%%%%%%%%%%%%%%%%%%%%%%%%%
%       Section          %
%%%%%%%%%%%%%%%%%%%%%%%%%%
\section{Generators of reduced dynamics and commutation relations}
\label{SecGenNLev}

Any operator that satisfies the hermitian \eqref{Gadjsym} and trace \eqref{Gtrace} condition, for example, the generator of reduced dynamics, can be written as a linear sum of $\i R_i, H_{ij}$ and $P_{ij}$, see \Eqs{Ri}-\eqref{Pij},
%%%
\begin{align}
    K&=\sum_{i=1}^N \w_i \,\i R_i+\sum_{i\leq j=1}^{N}\al_{ij}H_{ij}+\sum_{i<j=1}^{N}\bt_{ij}P_{ij}\,,
\end{align}
%%%
where $\w_i, \al_{ij}$ and $\bt_{ij}$ are real coefficients. To extract the coefficients for a given $K$, we first write the $\times$-product as a tensor product $a\times b\rightarrow X=a\otimes b^\t$, where $\t$ denotes the transpose of a matrix. Then, we define the trace norm $||X||\equiv \tr(a)\tr(b)$. We notice that $\i R_{i}, L_j, T_{ij}, \i A_{ij}$ and $I_N$ are mutually orthogonal under the norm $||X^\dg Y||$, which can be deduced by using \Eq{tt} and the fact that all of the $\lam$-matrices are traceless. The coefficients can then be extracted as follows,
%%%
\begin{subequations}
\begin{align}
    \w_i&=\frac{1}{N}||(\i R_i)^\dg K||\,,\\
    \al_{ii}&=\frac{1}{2}||T^\dg_{ii}K||\,,\\
    \al_{ij}&=||T^\dg_{ij}K||\,,\\
    \bt_{ij}&=||(\i A_{ij})^\dg K||\,.
\end{align}
\end{subequations}
%%%

The generators have the following commutation relations,
%%%
%\begin{widetext}
\begin{subequations}
\begin{align} \label{RRcomm}
    [\i R_i,\i R_j]&=- f_{ijk}\i R_k\,,\\
    [\i R_i,H_{mn}]&=f_{ir\u{m}}H_{\u{n}r}\,,\\
    [\i R_i,P_{mn}]&=-f_{ir\hat{m}}P_{\hat{n}r}\,,
\end{align}
%%%
%%%
\begin{align}
    &[H_{ij},H_{mn}]=\left(-\frac{2}{N}\del_{\u{i}\ud{m}}f_{\ud{n}\u{j}r}
                            +d_{ijk}d_{mns}f_{ksr}\right)\i R_r \nm
                        &\quad+d_{mns}f_{sr\u{i}} P_{\u{j}r}-d_{ijs}f_{sr\u{m}} P_{\u{n}r}
                        +d_{r\u{i}\ud{m}}f_{\ud{n}\u{j}s} P_{rs}\,,
\end{align}
%%%
%%%
\begin{align}
    [H_{ij},P_{mn}]&=d_{ijt}f_{mnr}f_{rst}\i R_s +d_{s\hat{m}\u{i}}f_{\u{j}\hat{n}r}H_{rs}\nm
                    &\quad-d_{ijs}f_{sr\hat{m}}H_{\hat{n}r} +f_{mnr}f_{rs\u{i}}P_{\u{j}s}\,,
\end{align}
%%%
%%%
\begin{align}
    &[P_{ij},P_{mn}]=\left(\frac{2}{N}\del_{\hat{i}\hhat{m}}f_{\hhat{n}\hat{j}r}
                            +f_{ijk}f_{mns}f_{ksr}\right)\i R_r \nm
                        &\quad+f_{mns}f_{sr\hat{i}}H_{\hat{j}r}-f_{ijs}f_{sr\hat{m}}H_{\hat{n}r}
                        -d_{r\hat{i}\hhat{m}}f_{\hhat{n}\hat{j}s} P_{rs}\,,\label{PPcomm}
\end{align}
\end{subequations}
%\end{widetext}
%%%
in which underlined and double-underlined indices should be symmetrized separately, whereas hatted and double-hatted indices should be anti-symmetrized separately.
For instance,
%%%
\begin{align}   \label{symanti-sym}
    d_{s\hat{m}\u{i}}f_{\u{j}\hat{n}r}&=d_{s\hat{m}i}f_{j\hat{n}r}+d_{s\hat{m}j}f_{i\hat{n}r}\nm
        &=d_{smi}f_{jnr}-d_{sni}f_{jmr}+d_{smj}f_{inr}-d_{snj}f_{imr}\,,
\end{align}
%%%
and so on.
To obtain \Eqs{RRcomm}-\eqref{PPcomm}, we have made use of the following identities \cite{MacFarlane68,MacFarlane01},
%%%
\begin{subequations}
\begin{align}   \label{iden1}
    d_{r(ns}f_{m)ir}&=d_{rns}f_{mir}+d_{rmn}f_{sir}+d_{rsm}f_{nir}=0\,,
\end{align}
%%%
%%%
\begin{align}
    f_{r(im}f_{n)sr}=0\,,\label{iden2}
\end{align}
%%%
%%%
\begin{align}
    f_{ijr}f_{mnr}=\frac{2}{N}(\del_{im}\del_{jn}-\del_{in}\del_{jm})+d_{imr}d_{jnr}-d_{inr}d_{jmr}\,,
\end{align}
\end{subequations}
%%%i
where round bracket over indices denotes a sum over the cyclic permutations of the indices under it, as shown explicitly in \Eq{iden1}.

Out of the generators in \Eqs{Ri}-\eqref{Pij}, the only one that fulfills the unitary condition \eqref{Gunitary} to generate compact transformation is $\i R_i$.
In fact, it produces ordinary unitary rotation with a factorized form,
%%%
\begin{align}   \label{expRt}
    e^{-\th \i R_i}&=e^{-\i \th \lam_i}\times e^{\i \th \lam_i}\,.
\end{align}
%%%
Therefore, it generates the ordinary symmetry $\rho'=u\rho u^\dg$, where $u\equiv e^{-\i \th \lam_i}$. Other general transformation which do not have this factorized form are not compact.

%\bibliography{GenSym}{}
\providecommand{\noopsort}[1]{}\providecommand{\singleletter}[1]{#1}%

\end{document}